\documentclass[a4paper,fleqn]{cas-sc}

\usepackage{aas}
\usepackage[numbers,sort&compress]{natbib}
\usepackage[latin9]{inputenc}
\usepackage{color}
\usepackage{amsmath}
\usepackage[colorinlistoftodos,prependcaption]{todonotes}
\presetkeys%
{todonotes}%
{inline,backgroundcolor=orange}{}
\tikzset{/tikz/notestyleraw/.append style={text=black!100!white}}

\makeatletter
\@ifundefined{textcolor}{}
{%
 \definecolor{BLACK}{gray}{0}
 \definecolor{WHITE}{gray}{1}
 \definecolor{RED}{rgb}{1,0,0}
 \definecolor{GREEN}{rgb}{0,1,0}
 \definecolor{BLUE}{rgb}{0,0,1}
 \definecolor{CYAN}{cmyk}{1,0,0,0}
 \definecolor{MAGENTA}{cmyk}{0,1,0,0}
 \definecolor{YELLOW}{cmyk}{0,0,1,0}
}

\usepackage{soul}

\usepackage{amsmath}
\usepackage{amsfonts}
\usepackage[nameinlink]{cleveref}
\makeatother

\def\tsc#1{\csdef{#1}{\textsc{\lowercase{#1}}\xspace}}
\tsc{WGM}
\tsc{QE}
\tsc{EP}
\tsc{PMS}
\tsc{BEC}
\tsc{DE}

\def\equationautorefname~#1\null{%
  Eq.~(#1)\null
}

\Crefname{table}{Table}{Tables}
\Crefname{figure}{Figure}{Figures}
\Crefname{section}{Sect.}{Sects.}
\Crefname{equation}{Eq.}{Eqs.}
\Crefname{eqnarray}{Eq.}{Eqs.}
\Crefname{appsec}{Appendix}{Appendices}

\begin{document}
\let\WriteBookmarks\relax
\def\floatpagepagefraction{1}
\def\textpagefraction{.001}
\shorttitle{Beyond self-acceleration}
\shortauthors{L.Amendola, V. Pettorino}
\title [mode = title]{Beyond self-acceleration: force- and fluid-acceleration}
\author[1]{Luca Amendola}
\fnmark[1]
\ead{l.amendola@thphys.uni-heidelberg.de}
\address[1]{Institut fuer Theoretische Physik, Ruprecht-Karls-Universitaet Heidelberg, Philosophenweg 16, 69120 Heidelberg, Germany}
\author[2]{Valeria Pettorino}
\fnmark[2]
\ead{valeria.pettorino@cea.fr}
\address[2]{AIM, CEA, CNRS, Universit\'e Paris-Saclay, Universit\'e Paris Diderot, Sorbonne Paris Cit\'e, F-91191 Gif-sur-Yvette, France}

\begin{abstract}
The notion of self acceleration has been introduced 
as a  convenient way to theoretically distinguish cosmological models in which acceleration is due to  modified gravity from those in which it is due to the properties of matter or fields. In this paper we 
 review the concept of self acceleration as given, for example, by \cite{2012PhRvL.109x1301W}, and highlight two problems. First, that it applies only to universal couplings, and second, that it is too narrow, i.e. it excludes models in which the acceleration can be shown to be induced by a genuine modification of gravity, for instance coupled dark energy with a universal coupling, the Hu-Sawicki $f(R)$ model or, in the context of inflation, the Starobinski model.
We then propose two new, more general, concepts in its place: force-acceleration and field-acceleration, which are also applicable in presence of non universal cosmologies. We illustrate their concrete application with two examples, among the modified gravity classes which are still in agreement with current data, i.e. $f(R)$ models and coupled dark energy.
\end{abstract}


\maketitle

\section{Introduction}

The origin of cosmic acceleration is one of the most
puzzling and debated questions in contemporary cosmology: the biggest
challenge relies in understanding whether acceleration is caused by a
cosmological constant or not. Alternatives to a cosmological constant
roughly divide into two main approaches: the first considers an additional
dynamical new fluid that adds to the usual cosmologically relevant
matter fields, i.e. dark matter, baryons, radiation, neutrinos; the
second, appealing path, tries to attribute acceleration to a change
in gravity and General Relativity itself (which typically also requires
new degrees of freedom associated with new fields). If a new field
is present, this could couple minimally (i.e. purely gravitationally,
as in the standard General Relativity) or non-minimally to the other
matter fields or to gravity or both. Moreover, the coupling could be
either universal or depend on the matter field, thereby violating
the equivalence principle. The notion of self acceleration was introduced after the discovery of cosmic acceleration to denote accelerated models without a cosmological constant (see e.g. \cite{2001PhLB..502..199D,2005PhRvD..72l3511K, 2009PhRvD..80b4037C, 2009PhRvD..80l1301S, 2011JCAP...04..025K}).
This classification was then generalized and made more precise in \cite{2012PhRvL.109x1301W}, and it is to this definition that we refer specifically in the following. Since then, it has often been used to highlight those cosmologies in which cosmic acceleration is due to a `pure' modification of gravity (see e.g. \cite{2016JCAP...03..031L,2017PhLB..765..382L, 2017JCAP...10..020R,2017PhRvD..96j4036D,2018PhRvD..97l4014N,2018PhRvD..97h4004C, 2018PhRvD..98d4048J}), and also employed  to motivate (or not) further investigation of certain classes of modified gravity models.  

In this paper we review what this concept actually means and discuss more precisely its limitations. We then propose a new classification that addresses such limitations, with the same original motivation in mind as for self acceleration, but applicable also to non universal couplings to matter species. This is particularly interesting in view of the fact that non universal coupled dark energy are among the classes of modified gravity models which are still compatible with all current observations: if baryons are minimally coupled to gravity, in fact,   local gravitational constraints are automatically satisfied,  as  well as the recent constraints on the speed of gravitational waves.

It is to be noted that the concept of self-acceleration is purely theoretical. Since all we can measure from cosmological observations in a homogeneous and isotropic universe is the expansion rate $H(z)$ and the space curvature, it is impossible to determine from cosmological observations alone whether our universe is self-accelerated or not. 
We will not discuss the issue of observability further in this paper and rather concentrate on the concept of self-acceleration as presented in literature, i.e. as a mean to determine whether, within a given theory, the predicted (theoretical) cosmic acceleration is to be attributed to a  modification of gravity or not.

\section{Reference frames}
In order to discuss self-acceleration, its shortcomings and possible extensions, it is necessary to review the concept of reference frames in some detail.

Observational data are in general obtained under the reasonable assumption that our
reference laboratory system (units of time, distance, mass) does not vary in time and space.
This means that the \emph{observed} reference system is the one in which
baryons are coupled to that metric for which this is true: we define then the
{\it observational metric}  the one in which baryons follow geodesics. We refer
to $S_{obs}$, as to the action in which this condition holds. 

Let's consider a model containing Einstein gravity, matter, and a scalar field.
If
$T_{\mu\nu}^{(b)}$ is the baryon Energy-Momentum Tensor (EMT), then
the observational metric (or frame) is the one in which the
EMT for baryons is conserved: 
\begin{equation}
T_{\,\nu;\mu}^{(b)\mu}=0\,\,\,\,\ ({\rm {observational\,\,\,frame})}
\label{conservedEMT}
\end{equation}
If there is only one matter component, or if all matter components are coupled  to the same metric, i.e. in the case of universal couplings, the frame in which
baryons follow geodesics (with metric $g_{\mu\nu}$) is also called
Jordan frame: in other words, for a universal coupling, the observational frame happens to be coincident with the Jordan frame, in which matter is conserved in a covariant way. For a universal coupling, it is often convenient to consider also a second relevant reference
frame: the one in which any additional scalar degree of freedom is only minimally
coupled to gravity through a factor $\sqrt{-g}$, ie. the Einstein
frame. Of course, one could define also mixed reference frames in which both baryons and field are not minimally coupled.

In any other metric non-trivially related
to $g_{\mu\nu}$, the EMT will, in general, not be conserved. For example, for a 
metric related to $g_{\mu\nu}$ via a conformal coupling, such that $\tilde{g}_{\mu\nu}=f(\phi)g_{\mu\nu}$, with $f$ a generic function of a scalar field $\phi$, 
the right hand side in \autoref{conservedEMT} does not vanish anymore and is  a function of $\phi$, so that  the EMT of each species  is no longer separately conserved\footnote{In the following, we will provide examples that are related to Einstein metric $g_{\mu\nu}$ via a conformal coupling; we note however that similar considerations can be extended to a general coupling metric $\hat g_{\mu\nu}=f(\phi,g_{\mu\nu})$, including the so-called disformal metric $\hat g_{\mu\nu}=C(\phi)g_{\mu\nu}+D(\phi)\phi_{\mu}\phi_{\nu}$.}: 
\begin{equation}
T_{\,\nu;\mu}^{(b)\mu}=-\frac{f_{,\phi}}{2f}T^{(b)}\phi_{,\nu}\,\,\, .
\end{equation}

Let's consider now the possibility that the coupling
is not universal, i.e. such that different matter species may couple to gravity with different strengths, mediated by a scalar field. For example, baryons may still be coupled only minimally
to gravity, while a non-minimal coupling may affect dark matter or
neutrinos. In this case, there are (at least) three relevant frames: one in
which only baryons are conserved, one in which only dark matter (or neutrinos) is
conserved, and one in which the scalar field is conserved, i.e. is minimally coupled to gravity. The first
one is the observational frame, the third one can be called the Einstein
frame; the second is, in general, neither Jordan nor Einstein, 
and we can refer to it as being the ``dark matter (or neutrinos) frame''. Here again, one could also equivalently define an infinity of other frames in which all three components are coupled.
Let's explicit now the actions corresponding to these three relevant frames.

For an Action that contains gravity, a scalar field,
baryons and dark matter one has the three main cases which can be
obtained one from another through a conformal transformation: 
\begin{align}
S_{E} & =\int d^{4}x\sqrt{-g}[R+L_{\phi}(g_{\mu\nu})+L_{b}(f_{b}(\phi)g_{\mu\nu})+L_{dm}(f_{dm}(\phi)g_{\mu\nu})]\,,\\
S_{dm} & =\int d^{4}x\sqrt{-\tilde{g}}[f_{g}(\phi)\tilde{R}+\tilde{L}_{\phi}(\tilde{g}_{\mu\nu})+\tilde{L}_{b}(\tilde{f}_{b}(\phi)\tilde{g}_{\mu\nu})+\tilde{L}_{dm}(\tilde{g}_{\mu\nu})]\,,\\
S_{obs} & =\int d^{4}x\sqrt{-\hat{g}}[\hat{f}_{g}(\phi)\hat{R}+\hat{L}_{\phi}(\hat{g}_{\mu\nu})+\hat{L}_{b}(\hat{g}_{\mu\nu})+\hat{L}_{dm}(\hat{f}_{dm}(\phi)\hat g_{\mu\nu})]\label{eq:sobs}
\end{align}
where $R,\tilde{R},\hat{R}$ are the curvature scalars  written
in terms of $g,\tilde{g},\hat{g}$, respectively, defined as
\begin{align}
\tilde{g}_{\mu\nu} & =f_{dm}(\phi)g_{\mu\nu}\,,\\
\hat{g}_{\mu\nu} & =\tilde{f}_{b}(\phi)\tilde g_{\mu\nu}
\end{align}
and the following relations hold,
\begin{align}
f_{g} & =f_{dm}^{-1}\,,\\
\tilde{f}_{b} & =f_{b}f_{dm}^{-1}\,,\\
\hat{f}_{dm} & =\tilde{f}_{b}^{-1}=f_{b}^{-1}f_{dm}\,,\\
\hat{f}_{g} & =f_{g}\tilde{f}_{b}^{-1}=f_{g}f_{b}^{-1}f_{dm}=f_b^{-1}
\end{align}
The metric $g$ is always defined as the one from
which the uncoupled Ricci scalar of General Relativity is built.
If $f_{dm}=f_{b}$ (universal coupling), then the observational frame
coincides with the $dm$ frame, which in this case is also the standard
Jordan frame. If $f_{b}=1$, then the observational frame coincides
with the Einstein frame. For any other choice, the three frames are different choices of reference frames, although they are physically equivalent and one can always pick the most convenient one for calculation purposes. We stress that $S_{E}=S_{dm}=S_{obs}$: they are just
the same Action written with different variables. In Appendix A we show this explicitly.

\section{The concept of self-acceleration and its shortcomings}
Although in agreement with observations, a cosmological
constant is affected by fine tuning and coincidence problems. One
needs to tune the initial conditions with very high precision in order
to justify the observed, very small non-zero value of the cosmological constant
$\Lambda$; in addition, if the whole contribution of dark energy
is entirely due to a cosmological constant, then its energy density
happens to be of the same order of magnitude of matter density just at
the present time, while it has been completely negligible in the past
and it will dominate in the future, over all other species. These
choices impact the naturalness of the model and lack a more fundamental
prediction of the value assumed by $\Lambda$. Most dark energy models
have then been introduced in order to address these issues, trying
to replace the cosmological constant with a fluid or a modification
of gravity. Although a combination of cosmological constant and modified
gravity would still be interesting to detect, an important question
and a desirable property for dark energy models is to understand whether
they are able to explain acceleration without the introduction of
any cosmological constant in the Lagrangian. More in general, it
would be interesting to explain acceleration without the action of
matter field, i.e. without fluid negative pressure, be it vacuum energy
or a non-standard kinetic term. This property is often embodied in
the condition of so-called \emph{self-acceleration} formalized by Wang et al. in \cite{2012PhRvL.109x1301W} (we refer to this paper as WHK in the following):
a model is self-accelerated if the model does not produce acceleration
in the frame where gravity is uncoupled (Einstein frame) while it
is accelerated in the observational frame. 
Or, as WHK say, 
``by self-acceleration, we mean accelerated expansion in the Jordan
frame, while the Einstein-frame expansion rate is not accelerating.
This is a sensible definition, for the lack of acceleration in the
Einstein frame - where the Einstein, and therefore the standard Friedmann,
equations hold - is equivalent to the lack of dark energy. In self-accelerating
theories, the observed (Jordan-frame) cosmic acceleration stems entirely
from the conformal transformation [...], i.e., a genuine modified gravity
effect.''

Note that the WHK condition of self-acceleration given above does not
really prevent the existence of a potential or even of a cosmological constant: it just requires
this term not to be sufficient for the observed present
acceleration. The idea is that if the model is not accelerated in
the Einstein frame, where gravity and matter are clearly separated
(minimally coupled), while it is so in the observational frame, then
the origin of the acceleration cannot be entirely due to vacuum energy
but has to come from the transformation of the metric that brings
the Einstein frame into the observational frame: in other words, it
has to be due to a modification of gravity rather than to the matter
sector. Whether one can really distinguish observationally between
these two cases is another question that will not be discussed here
except to notice that, obviously, if one allows for a completely arbitrary
matter sector (including arbitrary deviation from a perfect fluid
tensor) then no observation based purely on gravitational effects
can distinguish between an effect due to the right-hand side
of Einstein equations to an effect due to the left-hand side.

Although we agree that the declaration of principle that led to such definition is worth being pursued, we feel there are two main limitations with the current definition of self-acceleration.
First, this definition  can only be literally
applied to the case of universal coupling (as indeed carefully specified
in Ref. \cite{2012PhRvL.109x1301W}) since in the non-universal coupling case the observational
frame can coincide with the Einstein frame and the self-acceleration
condition would become contradictory. Second, even if we restrict
ourselves to a universal coupling, this definition excludes from ``self-accelerated cosmologies'' even some models in which the acceleration
in the Einstein frame is not due to vacuum energy (or negative pressure)
but rather to the effect of a coupling between scalar field and
matter, as we will show below.
That is, it might exclude also models in which acceleration is
actually driven by a modification of gravity, contrary to the  intended
scope.

Goal of this short note is to elaborate a more general definition of
self-acceleration that, while keeping the original motivation, addresses these two shortcomings. Our new proposal is valid also when the coupling is non universal, and we will show that in literature there are indeed already examples of non-universal self-accelerated models that  fulfil our new definition, i.e. in which acceleration is only a consequence of a modification of gravity. Moreover, we will show that there are also cases  with a universal coupling in which the acceleration is indeed driven by a modification of gravity and yet would be classified as non-self-accelerated according to WHK definition, but not according to ours.

\section{Force-acceleration and field-acceleration} \label{def:force&field}
Let's then go back to the original motivation for introducing the concept of self-acceleration. 
The reason why this distinction  proves useful is because
most proposals for modifying gravity are motivated by the observation
of acceleration, so it is important to characterize them according
to whether they really induce acceleration or just modify the acceleration
that is already driven by the uncoupled matter sector. In other words,
whether modifying gravity is really needed, at least for what concerns
acceleration.
Let's then assume that we would like to distinguish between models that explain acceleration with vacuum energy or negative pressure and models that explain it with a \emph{modification of gravity}, i.e. with the addition of a propagating degree of freedom that mediates an extra force (assumed to be a scalar field for definiteness). A first attempt to generalise the
definition of self-acceleration that extends also to non-universal
couplings might be: \emph{a model is self-accelerated if the universe expansion is accelerated
when the couplings of the scalar field (to matter or to gravity, universal
or not) are non-zero but is not accelerated when they are taken to
vanish.} Here, "coupling to gravity" means specifically the existence of a covariant term in the Lagrangian that includes the scalar field and second derivative tensors of the metric.\footnote{This applies to models with  single-metric gravity and a scalar field. In case of multi-metric theories, see e.g. \cite{deRham:2010ik,Hassan:2011zd}, the direct coupling between metrics also acts as a modification of gravity. Our classification in force- and fluid-accelerated models remains valid.}
Consider then a cosmological model with matter and a scalar field: we want to check whether this is self-accelerated according to this definition. First, we would switch off the couplings (i.e. take them to vanish): this means looking at the limit in which we are in a cosmology with standard gravity, plus a scalar field, plus matter. If acceleration is present, then this cannot be due to a modification of gravity, but rather to the field or
matter, having for example strongly negative pressure, and there would therefore be no self-acceleration. If instead there is no acceleration when the couplings are set to zero, while there is acceleration when one or more couplings are switched on, this
must be due to the new interactions (forces) rather than to the matter or field
content. This definition has the advantage that it makes no use of the Jordan frame vs Einstein frame transformation, which is ambiguous for non-universal couplings.

Once one recognises that the divide between vacuum-energy-driven
and modified-gravity-driven acceleration is actually a divide between
uncoupled models and coupled ones, the expression ``self-acceleration''
appears not sufficiently clear. We propose then that a more useful
distinction is between \emph{force-acceleration} and \emph{fluid-acceleration}.
The distinction is an operational one. A model is
\begin{itemize}
    \item \emph{force-accelerated}
when the acceleration in the observational frame becomes deceleration when
all the couplings between matter and scalar field or
between scalar field and gravity are switched off, i.e. 
    \begin{equation}
    f_{g}=f_{dm}=f_{b}=1 \,\, ;
    \end{equation}
\item \emph{fluid-accelerated} when
the acceleration in the observational frame persists in absence of such
couplings. 
\end{itemize}
In a sufficiently complicated model one can have periods
of force-acceleration and periods of fluid-acceleration. Here, for
simplicity, we will refer to the present-time cosmic acceleration (but also show one possible example that applies to inflation in \autoref{sect:fR}).
A model which is neither force- nor fluid-accelerated,
has no acceleration and is therefore uninteresting to describe present cosmology.

Let's apply the definitions above to some quick examples of different theories, to clarify their scope. Consider an Action without any interaction term between
the scalar field and matter or gravity. If any of the matter fluids
accelerates the expansion (e.g. some form of dark matter with negative
pressure), then we are in a fluid-accelerated
case. The same if the scalar field induces acceleration because of
its potential or a non-standard kinetic term. If instead there is
no acceleration in absence of couplings, but there is when they are
switched on, then we are in a force-accelerated case. 

Taking the Horndeski Action \cite{Horndeski:1974wa,Deffayet:2009mn} as another example, we see that the two functions $\alpha_M,\alpha_T$ (defined in \cite{Bellini:2014fua}) parameterize the coupling of the Horndeski field to gravity, i.e. to the second-derivative tensors $R$ and $R_{\mu\nu}$. When $\alpha_{M,T}=0$ the scalar field is minimally coupled to gravity but it possesses, in general, a non-canonical kinetic sector. If in this case the expansion is accelerated, then it is a case of fluid-acceleration. If the acceleration occurs only when either $\alpha_M$ or $\alpha_T$ differ from zero, then we are in presence of  force-acceleration.

In the next section we highlight more in detail how our new definition  applies to two of the classes of theories for dark energy and modified gravity that are still interesting at present, as both survive present constraints from cosmology and from gravitational waves.

\section{Example 1: $f(R)$ models} \label{sect:fR}
The first class we consider here is based on a $f(R)$ cosmology. This class is a subclass of more general scalar tensor theories, and can be redefined in terms of a scalar field that couples to gravity in a non-minimal way. Such cosmologies have been studied for a long time and constrained in many ways (e.g. \cite{2003IJMPD..12.1969C, 2004PhRvD..70d3528C, 2007PhRvL..98m1302A} among many others; see also \cite{Capozziello2008,DeFelice:2010aj} for a review). After many years they still offer a range of values that fits current data; their tensor equation, describing gravitational wave propagation, is also  in agreement with a speed of $c_T = 1$, and therefore survives gravitational wave detection  \cite{Monitor:2017mdv}. As they are equivalent to having a universal coupling to gravity for all matter fields, including baryons, they would typically require some screening mechanism active in high density regions, to guarantee that solar system constraints are also satisfied. Also, since the coupling to gravity is universal, the Jordan frame coincides with the observational frame. In this frame, this class of models can be described with the following Action (we choose units such that $16\pi G=1$):
\begin{equation}
S = \int d^{4}x\sqrt{-g}[f(R)+L_{dm}(g_{\mu\nu})+L_{b}(g_{\mu\nu})] \,\, .\label{eq:frorig}
\end{equation}
In order to apply our definition, we want to express this Action into the scalar tensor form of \autoref{eq:sobs}. As well known, under certain conditions, this can be done via a Legendre transformation \cite{2007PhRvD..75f7302F, Pettorino_Baccigalupi_2008, 2010LRR....13....3D}.
We define 
\begin{equation}
\phi \equiv df/dR \,\, ,
\end{equation} 
and we write the Legendre transform as
$\hat{f}(\phi) \equiv f-\phi R$.
By differentiation, since $df=\phi dR$, and $d\hat{f}  =df-Rd\phi-\phi dR =-Rd\phi$, one finds that $d\hat{f}/d\phi = - R$, and therefore $\hat{f}$ depends only on $\phi$ (and not on derivatives of $\phi$). We can then always redefine $f(R)$ in terms of a potential $V(\phi)$ such that:
\begin{equation}
f = \phi R + \hat{f}(\phi) \equiv \phi R-V(\phi) \,, \label{phiR}
\end{equation}
with 
\begin{equation}
V(\phi) = -f(R)+\phi R 
\, , \label{Vdef}
\end{equation}
which becomes a function of $\phi$ alone when expressing $R$ in terms of $\phi$ by inverting $\phi=df/dR$. Comparing \autoref{phiR} with \autoref{eq:sobs}, we then have that any action 
of the kind of \autoref{eq:frorig} can be rewritten as 
\begin{equation}
S=d^{4}x\sqrt{-g}[\phi R-V(\phi)+L_{dm}(g_{\mu\nu})+L_{b}(g_{\mu\nu})] \,\, ,\label{eq:fr-e-action}
\end{equation}
where now $f_{g}=\phi$ and where the potential is given by \autoref{Vdef}. 

Suppose now, for instance, that the model of interest within the class in \autoref{eq:frorig} is
\begin{equation}
f(R) \equiv R+\alpha R^{2}-\Lambda \,\, ; \label{fRexample1}
\end{equation} 
then $\phi = 1+2\alpha R$, \autoref{eq:fr-e-action} holds and
\begin{align}
V(\phi) & =-R-\alpha R^{2}+\Lambda+(1+2\alpha R)R\\
 & =\Lambda+\alpha R^{2}=\Lambda+\frac{(\phi-1)^{2}}{4\alpha} \,\, .
\end{align}
According to our definition above, the model \autoref{fRexample1} is fluid-accelerated if it is accelerated when the couplings (in this case the conformal factors coupling non-minimally to gravity) are switched off, i.e. if we rewrite the Action (\ref{eq:fr-e-action})
with $f_{g}=1$, that is as
\begin{equation}
S=d^{4}x\sqrt{-g}[R-V(\phi)+L_{dm}(g_{\mu\nu})+L_{b}(g_{\mu\nu})] \,\, .\label{eq:froff}
\end{equation}
When varying with respect to $\phi$, one gets $\phi=1$, so $V=\Lambda$ and
finally this Action is equivalent to a standard $\Lambda$CDM action.
The model \autoref{fRexample1} is therefore indeed fluid-accelerated, provided of course $\Lambda>0$ and  sufficiently large today. This simple example
helps clarifying the content of our definition: the presence
of $\Lambda$ in the observed Action \autoref{fRexample1} means obviously that there is
no further need of modified gravity to explain acceleration.

If instead we start from 
\begin{equation}
    f(R)=R+\alpha R^{2} \,\, , \label{fRexample2}
\end{equation}
(also known as Starobinski's inflation \cite{Starobinsky:1980te}), switching off the coupling term reduces the model to a pure matter model without acceleration, because $V(\phi=1)=0$. Since Starobinski's
model is accelerated (not today, but we are using it only as illustration),
then we should classify it as force-accelerated model, i.e. past inflationary acceleration in this model is really a pure gravitational effect. 

For more complicated forms of $f(R)$, (and in general for scalar-tensor theories or extended quintessence \cite{perrotta_extended_1999} with canonical kinetic term) the criterion for fluid-acceleration vs force-acceleration reduces (as expected) to simply checking whether $V(\phi)$ alone is capable of supporting acceleration, in absence of couplings as in  \autoref{eq:froff}. The general condition on $f(R)$ for this to occur can be derived as follows. In the form (\ref{eq:froff}), Einstein equations are obtained varying with respect to $\phi$. This implies
$\partial V/\partial\phi=0$, and therefore, using \autoref{Vdef} into \autoref{eq:froff} and considering $\phi$ and R as independent degrees of freedom, $R(\phi)=0$.
Inserting this condition in $V(\phi)$ (\autoref{Vdef}) we have immediately $V = - f(0)$.
Then, if $f(0)$ is a non-zero constant, the uncoupled version of the model is $\Lambda$CDM
and therefore fluid-accelerated (again, provided $\Lambda$ is large enough to overcome matter's deceleration), otherwise either is not accelerated
at all or is force-accelerated. So the Hu-Sawicki model \cite{2007PhRvD..76f4004H}, for instance, is such that $f(R=0) = 0$, i.e. it is decelerated in absence of couplings, but accelerated with coupling: it is then force-accelerated, according to our definition above. Indeed, one of the main motivations behind the Hu-Sawicki model was to satisfy the condition $f(R\to 0)\to 0$.

If now we apply WHK's classification, we find that neither model \autoref{fRexample1} nor \autoref{fRexample2} is  self-accelerated, since they both are accelerated also in the Einstein frame. Starobinski's model is however a clear case in which acceleration is driven entirely by a modification of gravity, and we believe it should be classified as such. 
More in general, any evolution such that $\phi={\rm const}$ (as in a typical de Sitter state) is not self-accelerated under WHK condition, since then the Jordan and Einstein frames coincide.
For instance, as we show in App. B, depending on the value of the parameters, the Hu-Sawicki model turns out to be WHK self-accelerated only near the present time, but not in the asymptotic future when it approaches a de Sitter solution.

The force-acceleration classification applies also when one considers a non-universal coupling. In fact, one can generalize the $f(R)$ models so that  baryons are decoupled, thereby escaping all local gravity constraints. For this, it is sufficient
to write in \autoref{eq:frorig} $L_{b}(f_{b}g_{\mu\nu})$ with
$f_{b}=f_{g}=\phi$. In this way, when we remove the coupling $f_{g}$
from the gravity sector via a conformal transformation, we remove
also the coupling $f_{b}$ from the baryonic sector. The frame becomes
the observational one, since in this frame the baryons are uncoupled,
and local gravity constraints are automatically bypassed. Switching off the coupling in the observational frame  brings us back to \autoref{eq:fr-e-action}, therefore the conclusions for the universally coupled $f(R)$ model apply in exactly the same way to the non-universal case. In this way, the reach of our classification is extended with respect to the WHK one.

\section{Example II: Coupled Quintessence with universal coupling} \label{example2:CQ}
In this second example we discuss another model that would be classified
as non self-accelerated by the WHK criterion, but as force-accelerated
according to ours. Since this model is actually accelerated due to the coupling
of matter to the scalar field, we believe the acceleration is ``a
genuine modified gravity effect'' and fulfils the rationale for
a self-acceleration. The model in the Einstein frame is 
\begin{equation}
S_{E}=\int d^{4}x\sqrt{-g}[R+L_{\phi}(g_{\mu\nu})+L_{b}(f_{b}(\phi)g_{\mu\nu})+L_{dm}(f_{dm}(\phi)g_{\mu\nu})] \,\, ,
\end{equation}
with
\begin{equation}
L_{\phi}(g_{\mu\nu})=-\frac{1}{2}\phi_{,\mu}\phi^{,\mu}-V_{0}e^{-\lambda\phi} \,\, ,
\label{def:CQlagrangian}
\end{equation}
where $V_0$ and $\lambda$ are constants. The coupling can be either universal or non universal, but here we consider only a universal one. Let us take  
\begin{equation}
f_{b}=f_{dm} \equiv f \equiv e^{2\beta\phi}
\end{equation}
where for simplicity we assume $\beta$ (also constant) and  $\lambda$ to be positive (but the results are completely general), and from now on we use a subscript $m$ to denote both baryons and dark matter. This class of models has been studied many times in the past, beginning with \cite{Wetterich:1994bg}.

The conservation equations are (see also App. C)
\begin{align}
\nabla_{\mu}^{(g)}T_{(\phi)\nu}^{\mu} & =\beta\phi_{,\nu}T_{(m)}\\
\nabla_{\mu}^{(g)}T_{(m)\nu}^{\mu} & =-\beta\phi_{,\nu}T_{(m)}
\end{align}
where $T_{(m)\nu}^{\mu}=T^{\mu(0)}_{(m)\nu}$ (see also \cite{Amendola_2000, Pettorino_Baccigalupi_2008} and references in \cite{Ade:2015rim}). The upshot \cite{Amendola_2000} is that the model leads to
asymptotic acceleration (i.e., to a global stable attractor) either for {\it a}) $\lambda^{2}<2$ when $\lambda(\beta+\lambda)<3$
or for {\it b}) $\beta>\lambda/2$ when $\lambda(\beta+\lambda)>3$, where $\lambda$ is the slope of  the potential in \autoref{def:CQlagrangian}. Neither case is an exact de Sitter Universe.

To apply now the WHK classification and ours, we need to derive the Jordan  frame, which in this case of universal coupling coincides with the observational one. Then the Action becomes
\begin{equation} \label{Sobs:cq}
S_{obs}=\int d^{4}x\sqrt{-\hat g}[F(\phi)\hat R+L_{\phi}(\hat g_{\mu\nu})+L_{m}(\hat g_{\mu\nu})]
\end{equation}
where
\begin{equation}\label{eq:f-coup}
      F(\phi)=e^{-2\beta\phi }
\end{equation}
   and
\begin{equation}
L_{\phi}(\hat g_{\mu\nu})=
-\frac{1}{2}(1-6\beta^2)F(\phi)\phi_{,\mu}\phi^{,\mu}
-V_{0}e^{-(\lambda+4\beta)\phi}
\end{equation}
In this frame, one finds that the conditions for asymptotic acceleration are just the same as in the Einstein frame (see for instance  \cite{amendola_dark_2010}, Chap. 9.2, where $Q=\beta$ and $\lambda$ is our $\lambda+4\beta$). This immediately shows that WHK's criterion leads to the puzzling result that such coupled models are  either not accelerated at all, or not self-accelerated, notwithstanding the fact that gravity is indeed modified with an arbitrarily high strength.

Let us apply now our classification. 
We begin with case {\it a}). If  $\lambda^{2}<2$, we have acceleration in the observational frame, and since we have acceleration also for $\beta=0$, we would classify the model as fluid-accelerated. If instead $\lambda^2>2$, we would classify the model as not accelerated at all. So for case {\it a}) our definition agrees with WHK.

In case {\it b}),  $\beta>\lambda/2$. If $\lambda^2>2$, while in absence of coupling we would again fall back in a decelerated case, there is instead acceleration in the observational frame when (and only when) the coupling to gravity $\beta$ is switched on. This case is therefore  for us a force-accelerated model. Acceleration is here induced by a large universal coupling $\beta>\sqrt{2}/2$ and is not present when $V(\phi)$ only is active, and therefore represents a clear case of modified gravity acceleration (with a universal coupling) that would have been missed if using the WHK definition for self-acceleration.

In Appendix B we express this model in the Horndeski form and further discuss our result in view of Ref. \cite{2017PhLB..765..382L}, which employs the WHK definition to exclude a vast class of models.

\section{Conclusions}
In this paper we have rediscussed the condition of self-acceleration, often used in literature to theoretically distinguish cosmological models in which acceleration is due to a pure modification of gravity from those where acceleration is the effect of the energy associated to a fluid. Such classification has been  employed to motivate further (or not) investigation of classes of models.
While keeping the same original motivation, as presented in \cite{2012PhRvL.109x1301W}, we highlighted, via examples, two limitations of the self-acceleration definition, as currently adopted in literature: first, self-acceleration cannot be directly applied to non-universal couplings; second, even in presence of a universal coupling, some models in which acceleration is clearly an effect of modified gravity would be considered as not self-accelerated.

We then proposed in \autoref{def:force&field} to reformulate the self-acceleration concept in two new, more general, definitions that address such limitations -- fluid-acceleration and field-acceleration -- and further discussed in a few examples how these definitions are applied in practice, with respect to self-acceleration. 

In particular, in \autoref{sect:fR} we discussed $f(R)$ (in various forms, including Hu-Sawicky) and the condition for this class to be fluid-accelerated or force-accelerated; we highlighted how this condition reduces, as expected, to testing whether the potential $V(\phi)$ is able to produce acceleration, once the theory has been expressed in the form \autoref{eq:froff} with non-minimal couplings switched off. Specifically, 
we find that  while  Starobinsky's $f(R)$ inflation is not self-accelerated according to WHK,  it should be classified as force-accelerated according to our definitions. Moreover, we find the Hu-Sawicki model to be always force-accelerated at the present time and in the future, while the WHK  criterion classifies it as  self-accelerated only near the present time and for $n>0.75$.
 In a second example (\autoref{example2:CQ}), we showed how the self-acceleration and our definitions lead again to different results: in particular, applying WHK's definition would exclude a range of models in presence of a universal coupling (class \emph{b)} in \autoref{Sobs:cq}, with $\beta > \lambda/2 > \sqrt{2}/2$), even when the acceleration is clearly due to a modification of gravity. 

As noted already for example in \cite{2017PhRvD..95h4029B, 2018A&A...619A..38P}, we further remark that at present non-universal couplings are among the (few) classes of models which survive gravitational wave detection and local constraints (see \cite{2018PhRvD..97h4004C} for a review on models surviving with a universal coupling). This is because, by construction,  baryonic interactions are  standard and satisfy solar system constraints; furthermore the speed of gravitational waves in these models is $c_T = 1$ and therefore in agreement with gravitational wave detection. It has also been noted (see for example \cite{Pettorino:2012ts,2013PhRvD..88f3519P, 2013PhRvD..88b3531S} and the update in \cite{Ade:2015rim}) that models in which a non-universal coupling between dark matter particles is considered would also solve the tension in the measurement of the Hubble parameter \cite{2019arXiv190307603R} due to the degeneracy $\beta-H_0$ first noted in Ref. \cite{Amendola:2003eq}.

\section*{Acknowledgments}
We thank Lam Hui, Dario Bettoni, Guillem Domenech, Lucas Lombriser, Marco Crisostomi, Christof Wetterich, and Filippo Vernizzi for useful discussions on some aspects of this project. Special thanks to Javier Rubio for discussions and for helping us with App. C.

  \crefalias{section*}{appsec}

\section*{Appendix A}
Here we explicitly write some relations holding for a conformal coupling transformation, which are used in the text and are convenient to keep at hand. We also explicitely recall and show that $S_{E}=S_{dm}=S_{obs}$: they are just the same Action written with different variables. 

If $R$ is the curvature written in terms of the Einstein metric $g_{\mu\nu}$, a conformal transformation of the type
\begin{equation}
\tilde{g}_{\mu\nu}  =f(\phi)g_{\mu\nu}\,,\\
\end{equation}
where $f$ is a generic function of $\phi$, leads to the following relations:
\begin{align}
\sqrt{-g} & =f^{-2}\sqrt{-\tilde{g}}\,,\\
R & =f(\tilde{R}+6\tilde{\Box}\omega-6\tilde{g}^{\mu\nu}\omega_{,\mu}\omega_{,\nu})\,,\\
\tilde{\Box} & =f^{-1}[2\omega_{,\mu}g^{\mu\nu}\partial_{\nu}+\Box] \,\, ,
\end{align}
where $\omega_{,\mu}\equiv f_{,\mu}/2f=f'\phi_{\mu}/(2f)$ and  $f'\equiv df/d\phi$, and $\tilde{R}$  is the curvature scalar written in terms of $\tilde{g}$. It follows that
\begin{align}
f^{-1}(R-6f\tilde{\Box}\omega_{,\mu}+6f\tilde{g}^{\mu\nu}\omega_{,\mu}\omega_{,\nu}) & =\tilde{R}\,,\\
f^{-1}(R-6\omega_{,\mu}g^{\mu\nu}\omega_{,\nu}-6\Box\omega) & =\tilde{R} \,\, .
\end{align}
Here we further give an explicit example of the conformal transformation procedure from $S_E$ to $S_{dm}$:
\begin{align}
S_{E} & =\int d^{4}x\sqrt{-g}[R+L_{\phi}(g_{\mu\nu})+L_{b}(f_{b}g_{\mu\nu})+L_{dm}(f_{dm}g_{\mu\nu})]\\
 & =\int d^{4}x\sqrt{-\tilde{g}}f_{dm}^{-2}[f_{dm}(\tilde{R}+6\tilde{\Box}\omega-6\tilde{g}^{\mu\nu}\omega_{,\mu}\omega_{,\nu})+L_{\phi}(g_{\mu\nu})+L_{b}(f_{b}g_{\mu\nu})+L_{dm}(\tilde{g}_{\mu\nu})]\\
 & =\int d^{4}x\sqrt{-\tilde{g}}\{f_{dm}^{-1}\tilde{R}+[f_{dm}^{-2}L_{\phi}(f_{dm}^{-1}\tilde{g}_{\mu\nu})+f_{dm}^{-1}(6\tilde{\Box}\omega-6\tilde{g}^{\mu\nu}\omega_{,\mu}\omega_{,\nu})] + \nonumber \\ &\,\,\,\,\,\,\,\,\,\, + f_{dm}^{-2}L_{b}(f_{b}f_{dm}^{-1}\tilde{g}_{\mu\nu})+f_{dm}^{-2}L_{dm}(\tilde{g}_{\mu\nu})\}\\
 & =\int d^{4}x\sqrt{-\tilde{g}}\{f_{dm}^{-1}\tilde{R}+\tilde{L}_{\phi}(\tilde{g}_{\mu\nu})+\tilde{L}_{b}(f_{b}f_{dm}^{-1}\tilde{g}_{\mu\nu})+\tilde{L}_{dm}(\tilde{g}_{\mu\nu})\}=S_{dm}
\end{align}
where $\tilde{L}_{b,dm}=f_{dm}^{-2}L_{b,dm}$, while the scalar Lagrangian
is
\begin{align}
\tilde{L}_{\phi}(\tilde{g}_{\mu\nu}) & =f_{dm}^{-2}L_{\phi}(f_{dm}^{-1}\tilde{g}_{\mu\nu})+f_{dm}^{-1}(6\tilde{\Box}\omega-6\tilde{g}^{\mu\nu}\omega_{,\mu}\omega_{,\nu})
\end{align}
If 
\begin{equation}
L_{\phi}=-\frac{1}{2}\phi_{,\mu}g^{\mu\nu}\phi_{,\nu}-V(\phi)
\end{equation}
then (here $f\equiv f_{dm}$ )
\begin{align}
\tilde{L}_{\phi}(\tilde{g}_{\mu\nu}) & =f^{-2}L_{\phi}(f^{-1}\tilde{g}_{\mu\nu})+f^{-1}(6\tilde{\Box}\omega-6\tilde{g}^{\mu\nu}\omega_{,\mu}\omega_{,\nu})\\
 & =-\frac{1}{2}\phi_{,\mu}f^{-1}\tilde{g}^{\mu\nu}\phi_{,\nu}-f^{-2}V(\phi)+f^{-1}(6\tilde{\Box}\omega-6\tilde{g}^{\mu\nu}\omega_{,\mu}\omega_{,\nu})\\
 & =-\frac{1}{2}\phi_{,\mu}f^{-1}\tilde{g}^{\mu\nu}\phi_{,\nu}-f^{-2}V(\phi)+6f^{-2}f'\phi_{,\mu}\tilde{g}^{\mu\nu}\omega_{,\nu}-6f^{-1}\tilde{g}^{\mu\nu}\omega_{,\mu}\omega_{,\nu}\\
 & =f^{-1}\tilde{g}^{\mu\nu}(-\frac{1}{2}\phi_{,\mu}\phi_{,\nu}+6f^{-1}f'\phi_{,\mu}\omega_{,\nu}-6\omega_{,\mu}\omega_{,\nu})-f^{-2}V(\phi)\\
 & =f^{-1}\tilde{g}^{\mu\nu}(-\frac{1}{2}\phi_{,\mu}\phi_{,\nu}+3\left(\frac{f'}{f}\right)^{2}\phi_{,\mu}\phi_{,\nu}-6\left(\frac{f'}{2f}\right)^{2}\phi_{,\mu}\phi_{,\nu})-f^{-2}V(\phi)\\
 & =-\frac{1}{2}f^{-1}\tilde{g}^{\mu\nu}[1-3\left(\frac{f'}{f}\right)^{2}]\phi_{,\mu}\phi_{,\nu}-f^{-2}V(\phi)\label{eq:finlpsi}
\end{align}
(in the third line we used integration by parts) which can be written in fully canonical form defining a new field
and a new potential 
\begin{align}
\psi_{,\mu} & =f^{-1/2}\left[1-3\left(\frac{f'}{f}\right)^{2}\right]^{1/2}\phi_{,\mu}\, \, ,\\
U(\psi) & =\frac{V(\phi)}{f^{2}(\phi)}\big|_{\phi(\psi)} \,\, .
\end{align}

Similarly, one can show that $S_{dm}=S_{obs}$ as follows.
\begin{align}
S_{dm} & =\int d^{4}x\sqrt{-\tilde{g}}[f_{g}(\phi)\tilde{R}+\tilde{L}_{\phi}(\tilde{g}_{\mu\nu})+\tilde{L}_{b}(\tilde{f}_{b}(\phi)\tilde{g}_{\mu\nu})+\tilde{L}_{dm}(\tilde{g}_{\mu\nu})]\\
 & =\int d^{4}x\sqrt{-\hat{g}}\tilde{f}_{b}^{-2}[f_{g}\tilde{f}_{b}(\hat{R}+6\hat{\Box}\omega-6\hat{g}^{\mu\nu}\omega_{,\mu}\omega_{,\nu})+\tilde{L}_{\phi}(\tilde{f}_{b}^{-1}\hat{g}_{\mu\nu})+\tilde{L}_{b}(\hat{g}_{\mu\nu})+\tilde{L}_{dm}(\tilde{f}_{b}^{-1}\hat{g}_{\mu\nu})]\\
 & =\int d^{4}x\sqrt{-\tilde{g}}\{\tilde{f}_{b}^{-1}f_{g}(\hat{R}+6\hat{\Box}\omega-6\hat{g}^{\mu\nu}\omega_{,\mu}\omega_{,\nu})+\tilde{f}_{b}^{-2}\tilde{L}_{\phi}(\tilde{f}_{b}^{-1}\hat{g}_{\mu\nu})+\tilde{f}_{b}^{-2}\tilde{L}_{b}(\hat{g}_{\mu\nu})+\tilde{f}_{b}^{-2}\tilde{L}_{dm}(\tilde{f}_{b}^{-1}\hat{g}_{\mu\nu})\}\\
 & =\int d^{4}x\sqrt{-\tilde{g}}\{\tilde{f}_{b}^{-1}f_{g}(\hat{R})+[\tilde{f}_{b}^{-2}\tilde{L}_{\phi}(\tilde{f}_{b}^{-1}\hat{g}_{\mu\nu})+\tilde{f}_{b}^{-1}f_{g}(6\hat{\Box}\omega-6\hat{g}^{\mu\nu}\omega_{,\mu}\omega_{,\nu})]+\tilde{f}_{b}^{-2}\tilde{L}_{b}(\hat{g}_{\mu\nu})+\tilde{f}_{b}^{-2}\tilde{L}_{dm}(\tilde{f}_{b}^{-1}\hat{g}_{\mu\nu})\}\\
 & =\int d^{4}x\sqrt{-\tilde{g}}\{\hat{f}_{g}\hat{R}+[\tilde{f}_{b}^{-2}\tilde{L}_{\phi}(\tilde{f}_{b}^{-1}\hat{g}_{\mu\nu})+\tilde{f}_{b}^{-1}f_{g}(6\hat{\Box}\omega-6\hat{g}^{\mu\nu}\omega_{,\mu}\omega_{,\nu})]+\hat{L}_{b}(\hat{g}_{\mu\nu})+\hat{L}_{dm}(\hat{f}_{dm}\hat{g}_{\mu\nu})\}\\
 & =\int d^{4}x\sqrt{-\tilde{g}}\{\hat{f}_{g}\hat{R}+\hat{L}_{\phi}(\hat{g}_{\mu\nu})+\hat{L}_{b}(\hat{g}_{\mu\nu})+\hat{L}_{dm}(\hat{f}_{dm}\hat{g}_{\mu\nu})\}=S_{obs}
\end{align}
where $\hat{L}_{b,dm}=f_{b}^{-2}\tilde{L}_{b,dm}$, while the scalar
Lagrangian is
\begin{align}
\hat{L}_{\phi}(\hat{g}_{\mu\nu}) & =\tilde{f}_{b}^{-2}\tilde{L}_{\phi}(\tilde{f}_{b}^{-1}\hat{g}_{\mu\nu})+\tilde{f}_{b}^{-1}f_{g}(6\hat{\Box}\omega-6\hat{g}^{\mu\nu}\omega_{,\mu}\omega_{,\nu})
\end{align}
If now we take $\tilde{L}_{\phi}$ in \autoref{eq:finlpsi}, written in the canonical form for
the rescaled field $\psi,$
\begin{equation}
\tilde{L}_{\phi}=-\frac{1}{2}\psi_{,\mu}\tilde{g}^{\mu\nu}\psi_{,\nu}-U(\psi)=-\frac{1}{2}\tilde{f}_{b}\psi_{,\mu}\hat{g}^{\mu\nu}\psi_{,\nu}-U(\psi)
\end{equation}
then 
\begin{align}
\hat{L}_{\phi}(\hat{g}_{\mu\nu}) & =\tilde{f}_{b}^{-2}\tilde{L}_{\phi}(\tilde{f}_{b}^{-1}\hat{g}_{\mu\nu})+\tilde{f}_{b}^{-1}f_{g}(6\hat{\Box}\omega-6\hat{g}^{\mu\nu}\omega_{,\mu}\omega_{,\nu})\\
 & =-\tilde{f}_{b}^{-1}\frac{1}{2}\psi_{,\mu}\hat{g}^{\mu\nu}\psi_{,\nu}-\tilde{f}_{b}^{-2}U(\psi)+\tilde{f}_{b}^{-1}f_{g}(6\hat{\Box}\omega-6\hat{g}^{\mu\nu}\omega_{,\mu}\omega_{,\nu})\\
 & =-\frac{1}{2}\psi_{,\mu}\tilde{f}_{b}^{-1}\hat{g}^{\mu\nu}\psi_{,\nu}-\tilde{f}_{b}^{-2}U(\psi)-6(\tilde{f}_{b}^{-1}f_{g})'\psi_{,\mu}\hat{g}^{\mu\nu}\omega_{,\nu}-6\tilde{f}_{b}^{-1}f_{g}\hat{g}^{\mu\nu}\omega_{,\mu}\omega_{,\nu}\\
 & =-\frac{1}{2}\psi_{,\mu}\tilde{f}_{b}^{-1}\hat{g}^{\mu\nu}\psi_{,\nu}-\tilde{f}_{b}^{-2}U(\psi)-6(\tilde{f}_{b}^{-1}f_{g})'\psi_{,\mu}\hat{g}^{\mu\nu}\frac{\tilde{f}_{b}'}{2\tilde{f}_{b}}\psi_{,\nu}-6\tilde{f}_{b}^{-1}f_{g}\hat{g}^{\mu\nu}(\frac{\tilde{f}_{b}'}{2\tilde{f}_{b}})^{2}\psi_{,\mu}\psi_{,\nu}\\
 & =\psi_{,\mu}\psi_{,\nu}[-\frac{1}{2}\tilde{f}_{b}^{-1}\hat{g}^{\mu\nu}-6(\tilde{f}_{b}^{-1}f_{g})'\hat{g}^{\mu\nu}\frac{\tilde{f}_{b}'}{2\tilde{f}_{b}}-6\tilde{f}_{b}^{-1}f_{g}\hat{g}^{\mu\nu}(\frac{\tilde{f}_{b}'}{2\tilde{f}_{b}})^{2}]-\tilde{f}_{b}^{-2}U(\psi)\\
 & =-\frac{1}{2}\tilde{f}_{b}^{-1}\psi_{,\mu}\psi_{,\nu}\hat{g}^{\mu\nu}\left[1-3(f_{g}-2\tilde{f}_{b}\frac{f'_{g}}{\tilde{f}_{b}^{'}})\left(\frac{\tilde{f}_{b}'}{\tilde{f}_{b}}\right)^{2}\right]-\tilde{f}_{b}^{-2}U(\psi)
\end{align}
where the prime here is derivative with respect to $\psi$. As earlier, this can
be written in fully canonical form defining a new field and a new
potential 
\begin{align}
\chi_{,\mu} & =\tilde{f}_{b}^{-1/2}[1-3(f_{g}-2\tilde{f}_{b}\frac{f'_{g}}{\tilde{f}_{b}^{'}})\left(\frac{\tilde{f}_{b}'}{\tilde{f}_{b}}\right)^{2}]^{1/2}\psi_{,\mu} \,\, ,\\
W(\chi) & =\frac{U(\psi)}{\tilde{f}_{b}^{2}}|_{\psi(\chi)} \,\, .
\end{align}

\section*{Appendix B}
 In Ref. \cite{2017PhLB..765..382L}  it was shown that WHK's self-acceleration is equivalent to the following condition for Horndeski models  with $\alpha_T=0$:
\begin{equation}\label{eq:lomb-cond}
    (1+\frac{H'}{H})(1+\frac{\alpha_M}{2})+\frac{\alpha_M'}{2}\le0
\end{equation}
($H$ is the Hubble function in the Jordan frame and prime is $d/d\ln a$)
while at the same time $1+\frac{H'}{H}>0$.
In coupled quintessence we have $\alpha_T=0$ and
\begin{equation}
    \alpha_M=\frac{F'}{F} =-2\beta\phi'
\end{equation}
where $F(\phi)$ is given in \autoref{eq:f-coup}
(notice that in Horndeski notation, $F=G_4$, see e.g. \cite{Bellini:2014fua}). Now, on any asymptotic acceleration solution, or more in general on any critical point of the coupled quintessence phase-space system (see \cite{amendola_dark_2010}), $\phi'=const$, so $\alpha_M=const$, and therefore  condition \autoref{eq:lomb-cond} reduces to
\begin{equation}\label{eq:lombr1}
    1+\frac{\alpha_M}{2}=1-\beta\phi' \le 0
\end{equation}
For case {\it a}), $\phi'=\frac{\lambda}{1+\beta\lambda}$, and one sees immediately that condition \autoref{eq:lombr1} is never satisfied (we are assuming $\beta,\lambda\ge 0$). For case {\it b}), one has instead
\begin{equation}
    \phi'=\frac{3}{\lambda+4\beta}
\end{equation}
and the self-acceleration condition becomes
\begin{equation}\label{eq:lombr2}
    1-\frac{3\beta}{\lambda+4\beta} \le 0
\end{equation}
It is not difficult to see that this condition cannot be realized since it contradicts the existence of case {\it b}), namely $\lambda(\lambda+\beta)>3$. Therefore, also case {\it b}) is always not self-accelerated, according to WHK and \cite{2017PhLB..765..382L}, notwithstanding the fact that acceleration only arises because $\beta$ is large. More in general, while according to Ref. \cite{2017PhLB..765..382L} values $\alpha_M>-2$ (assuming constant $\alpha_M$) are to be excluded because non self-accelerating, they might be acceptable according to our definition of force-acceleration (for $\beta\gg \lambda$, for instance, we have $\alpha_M=-3/2$).

One can use condition \autoref{eq:lomb-cond} also to confirm that Starobinski's $f(R)$ model, in the context of inflation, is not self-accelerated if one applies WHK condition. 
In fact, one can see from \autoref{eq:fr-e-action} that the Horndeski $\alpha_M$ parameter for $f(R)$ models is simply
\begin{equation}\label{eq:amfr}
    \alpha_M=\frac{\phi'}{\phi}
    =\frac{f_{,RR}}{f_{,R}}R' 
\end{equation}
where the comma denotes the derivative with respect to $R$ and $\phi=f_{,R}$.
In Starobinski's model during inflation, i.e. for $R\approx 12 H^2 \gg 1/\alpha$, one has $H\approx H_i-(t-t_i)/(36\alpha)$ (see e.g. \cite{DeFelice:2010aj}). It follows
\begin{equation}
    \alpha_M=\frac{\phi'}{\phi}\approx \frac{R'}{R}\approx\frac{2H'}{H}\approx -\frac{1}{18\alpha H^2}
\end{equation}
which is much less than unity during inflation, 
while $\alpha_M'\sim \alpha_M^2$ is even smaller, and therefore    \autoref{eq:lomb-cond} is not satisfied.

The same analysis can be extended to the Hu-Sawicki model, for which\footnote{Notice that our $f(R)$ is $R+f(R)$ in the notation of \cite{2007PhRvD..76f4004H}.}
\begin{equation}
    f(R)=R-m^2\frac{c_1(R/m^2)^n}{c_2(R/m^2)^n+1}
\end{equation}
where $n,m,c_1,c_2$ are positive constants.
According to \cite{2007PhRvD..76f4004H}, near the present epoch and in the future one can approximate the model 
as
\begin{equation}
    f\approx R-A+BR^{-n}
\end{equation}
with $A,B$  positive constants formed out of $m,c_1,c_2$. In this regime  one has \cite{2007PhRvD..76f4004H}
\begin{equation}
    R\approx 3m^2(a^{-3}+4\frac{1-\tilde\Omega_m}{\tilde\Omega_m})
\end{equation}
where $a$ is the cosmic scale factor normalized to unity today and $\tilde\Omega_m$ approximates the true present matter density fraction.
Then we have
\begin{equation}
    \alpha_M\approx \frac{3(n+1)}{1+4a^3\frac{1-\tilde\Omega_m}{\tilde\Omega_m}}
\end{equation}
Now the condition \autoref{eq:lomb-cond} can be written as
\begin{equation}
    S(a) \equiv -(\frac{1}{2}+\frac{3}{2}w_{DE}\Omega_{DE})(1+\frac{\alpha_M}{2})+\frac{\alpha_M'}{2}\le0
\end{equation}
where  $w_{DE},\Omega_{DE}$ are the equation of state and matter density parameter, respectively. For $\tilde\Omega_m=0.3$ and $w_{DE}=-1$ (since the Hu-Sawicki model approximates a $\Lambda$CDM expansion), one has today $S_0\approx 0.24-0.31n$, which is indeed negative for any   $n>0.75$. However, $S(a)$ becomes (and remains) positive at some point in the future: for instance, at $a=1.23$ for $n=2$ and at $a=1.5$ for $n=4$. In other words, applying the WHK definition, the Hu-Sawicki models turns out to be self-accelerated only near the present time. For an overview on constraints by current data see for example \cite{PhysRevD.90.103512, f29fd58f57684c3c9caabb4c1c45bb21, doi:10.1002/andp.201400058}.

\section*{Appendix C}\label{appendixA}
In the main text we discussed only the case of a pure conformal coupling. Here  we write down the explicit form of the coupling for a more general  scalar-scalar interaction (but still without derivative couplings). Consider two scalar fields $\phi$ and $\chi$ with Lagrangian
density 
\begin{equation}\label{action-a}
\frac{{\cal L}}{\sqrt{-g}}=R+L_{\phi}+L_{\chi}\,,
\end{equation}
where
\begin{eqnarray}
    L_{\phi}&=&-\frac{1}{2}\partial^{\mu}\phi\partial_{\mu}\phi-V(\phi)\,,\\
    L_{\chi}&=&-\frac{1}{2}K(\phi)\partial^{\mu}\chi\partial_{\mu}\chi-I(\phi,\chi)\,,
\end{eqnarray}
with  $V(\phi)$ a 
potential and $I(\phi,\chi)$ an interaction term. We obtain the following equations of motion 
\begin{equation}
-\nabla^{\nu}(\partial_{\nu}\phi)+V_{,\phi}=-\frac{\partial I}{\partial\phi}-\frac{1}{2}K_{,\phi}\,\partial^{\mu}\chi\partial_{\mu}\chi\,,\hspace{15mm}-K^{-1}\nabla^{\nu}(K\partial_{\nu}\chi)=-K^{-1}\frac{\partial I}{\partial\chi}\,.
\end{equation}
We define now the energy-momentum tensors
\begin{equation}
    T^{\phi,\chi}_{\mu\nu}=-\frac{2}{\sqrt{-g}}
    \frac{\partial\sqrt{-g}L_{(\phi,\chi)}}{\partial g^{\mu\nu}}\,.
\end{equation}
While the total energy momentum tensor $T_{\mu\nu}=T_{\mu\nu}^{\phi}+T_{\mu\nu}^{\chi}$
is conserved, the individual components $T_{\mu\nu}^{\phi}$ and $T_{\mu\nu}^{\chi}$
are not.
We find the conservation equations
\begin{align}
T_{(\chi)\nu;\mu}^{\mu} & =-\frac{1}{3}\left((-4\beta+2\gamma)T_{(\chi)\nu}^{\mu}\phi_{,\mu}+(\beta+\gamma)T_{(\chi)}\phi_{,\nu}\right)\,,\\
T_{(\phi)\nu;\mu}^{\mu} & =\frac{1}{3}\left((-4\beta+2\gamma)T_{(\chi)\nu}^{\mu}\phi_{,\mu}+(\beta+\gamma)T_{(\chi)}\phi_{,\nu}\right)\,,
\end{align}
with 
\begin{equation}
\beta\equiv-\frac{1}{2}\frac{\partial\ln I(\phi,\chi)}{\partial\phi}\,,\hspace{10mm}\gamma\equiv-\frac{1}{2}\frac{\partial\ln K(\phi)}{\partial\phi}\,.\label{betadef}
\end{equation}
When $K(\phi)=1$ and for a separable interaction term $I(\phi,\chi)=I_{1}(\phi)I_{2}(\chi)$,
the effective couplings \autoref{betadef} become $\chi$-independent.
In particular, for $I=m^{2}(\phi)\chi^{2}/2$, we have 
\begin{equation}
\beta(\phi)=-\frac{\partial\ln m(\phi)}{\partial\phi}\,.\label{betadef1}
\end{equation}
If the Action \autoref{action-a} has been obtained through a conformal transformation of an Action with two non-interacting bosons $\phi,\psi$ in which $\phi$ couples conformally to gravity, then $I$ is separable and $\gamma=2\beta$
(because the kinetic term has an extra $g^{\mu\nu}$ term with respect to the
potential term) and only the coupling $\pm\beta T_{(\chi)}\phi_{,\nu}$
is left in the conservation equations.
A similar procedure for a fermion-boson coupling leads to the same result.

\bibliographystyle{elsarticle-num}
\bibliography{varia,self-acc,refs}

\end{document}